\documentclass[12pt,a4paper]{article}
\usepackage{psfrag}
\usepackage{graphics}
\usepackage{amssymb,epsfig,amsmath,euscript,array}
\makeatletter
\@addtoreset{equation}{section}
\makeatother

\newcounter{multieqs}

\newcommand{\be}{\begin{equation}}
\newcommand{\ee}{\end{equation}}

\newcommand{\bm}[1]{\mbox{\boldmath $#1$}}

\def\bd{\begin{document}}
\def\ed{\end{document}}
\def\nn{\nonumber}
\def\bea{\begin{eqnarray}}
\def\eea{\end{eqnarray}}
\let\bm=\bibitem
\let\la=\label

\def\npb#1#2#3{Nucl. Phys. {\bf{B#1}} #3 (#2)}
\def\plb#1#2#3{Phys. Lett. {\bf{#1B}} #3 (#2)}
\def\prl#1#2#3{Phys. Rev. Lett. {\bf{#1}} #3 (#2)}
\def\prd#1#2#3{Phys. Rev. {D \bf{#1}} #3 (#2)}
\def\cmp#1#2#3{Comm. Math. Phys. {\bf{#1}} #3 (#2)}
\def\cqg#1#2#3{Class. Quantum Grav. {\bf{#1}} #3 (#2)}
\def\nppsa#1#2#3{Nucl. Phys. B (Proc. Suppl.) {\bf{#1A}}#3 (#2)}
\def\ap#1#2#3{Ann. of Phys. {\bf{#1}} #3 (#2)}
\def\ijmp#1#2#3{Int. J. Mod. Phys. {\bf{A#1}} #3 (#2)}
\def\rmp#1#2#3{Rev. Mod. Phys. {\bf{#1}} #3 (#2)}
\def\mpla#1#2#3{Mod. Phys. Lett. {\bf A#1} #3 (#2)}
\def\jhep#1#2#3{J. High Energy Phys. {\bf #1} #3 (#2)}
\def\atmp#1#2#3{Adv. Theor. Math. Phys. {\bf #1} #3 (#2)}

\newcommand{\EQ}[1]{\begin{equation} #1 \end{equation}}
\newcommand{\AL}[1]{\begin{subequations}\begin{align} #1 \end{align}\end{subequations}}
\newcommand{\SP}[1]{\begin{equation}\begin{split} #1 \end{split}\end{equation}}
\newcommand{\ALAT}[2]{\begin{subequations}\begin{alignat}{#1} #2 \end{alignat}\end{subequations}}
\def\beqa{\begin{eqnarray}}
\def\eeqa{\end{eqnarray}}
\def\beq{\begin{equation}}
\def\eeq{\end{equation}}

\def\N{{\cal N}}
\def\sst{\scriptscriptstyle}
\def\thetabar{\bar\theta}
\def\Tr{{\rm Tr}}
\def\one{\mbox{1 \kern-.59em {\rm l}}}
 \def\Nh{\hat{N}}

\def\a{\alpha}      \def\da{{\dot\alpha}}
\def\b{\beta}       \def\db{{\dot\beta}}
\def\c{\gamma}  \def\G{\Gamma}  \def\cdt{\dot\gamma}
\def\d{\delta}  \def\D{\Delta}  \def\ddt{\dot\delta}
\def\e{\epsilon}        \def\vare{\varepsilon}
\def\f{\phi}    \def\F{\Phi}    \def\vvf{\f}
\def\h{\eta}
\def\k{\kappa}
\def\l{\lambda} \def\L{\Lambda}
\def\m{\mu} \def\n{\nu}
\def\o{\omega}
\def\p{\pi} \def\P{\Pi}
\def\r{\rho}
\def\s{\sigma}  \def\S{\Sigma}
\def\t{\tau}
\def\th{\theta} \def\Th{\Theta} \def\vth{\vartheta}
\def\X{\Xeta}
\def\z{\zeta}

\def\cA{{\cal A}} \def\cB{{\cal B}} \def\cC{{\cal C}}
\def\cD{{\cal D}} \def\cE{{\cal E}} \def\cF{{\cal F}}
\def\cG{{\cal G}} \def\cH{{\cal H}} \def\cI{{\cal I}}
\def\cJ{{\cal J}} \def\cK{{\cal K}} \def\cL{{\cal L}}
\def\cM{{\cal M}} \def\cN{{\cal N}} \def\cO{{\cal O}}
\def\cP{{\cal P}} \def\cQ{{\cal Q}} \def\cR{{\cal R}}
\def\cS{{\cal S}} \def\cT{{\cal T}} \def\cU{{\cal U}}
\def\cV{{\cal V}} \def\cW{{\cal W}} \def\cX{{\cal X}}
\def\cY{{\cal Y}} \def\cZ{{\cal Z}}

\def\ua{\underline{\alpha}}
\def\ub{\underline{\phantom{\alpha}}\!\!\!\beta}
\def\uc{\underline{\phantom{\alpha}}\!\!\!\gamma}
\def\um{\underline{\mu}}
\def\ud{\underline\delta}
\def\ue{\underline\epsilon}
\def\una{\underline a}\def\unA{\underline A}
\def\unb{\underline b}\def\unB{\underline B}
\def\unc{\underline c}\def\unC{\underline C}
\def\und{\underline d}\def\unD{\underline D}
\def\une{\underline e}\def\unE{\underline E}
\def\unf{\underline{\phantom{e}}\!\!\!\! f}\def\unF{\underline F}
\def\unm{\underline m}\def\unM{\underline M}
\def\unn{\underline n}\def\unN{\underline N}
\def\unp{\underline{\phantom{a}}\!\!\! p}\def\unP{\underline P}
\def\unq{\underline{\phantom{a}}\!\!\! q}
\def\unQ{\underline{\phantom{A}}\!\!\!\! Q}
\def\unH{\underline{H}}

\def\As {{A \hspace{-6.4pt} \slash}\;}
\def\bs {{b \hspace{-6.4pt} \slash}\;}
\def\Ds {{D \hspace{-6.4pt} \slash}\;}
\def\ds {{\del \hspace{-6.4pt} \slash}\;}
\def\ss {{\s \hspace{-6.4pt} \slash}\;}
\def\ks {{ k \hspace{-6.4pt} \slash}\;}
\def\ps {{p \hspace{-6.4pt} \slash}\;}
\def\pas {{{p_1} \hspace{-6.4pt} \slash}\;}
\def\pbs {{{p_2} \hspace{-6.4pt} \slash}\;}

\def\Fh{\hat{F}}
\def\Vh{\hat{V}}
\def\Xh{\hat{X}}
\def\ah{\hat{a}}
\def\xh{\hat{x}}
\def\yh{\hat{y}}
\def\ph{\hat{p}}
\def\xih{\hat{\xi}}

\def\psit{\tilde{\psi}}
\def\Psit{\tilde{\Psi}}
\def\tht{\tilde{\th}}

\def\At{\tilde{A}}
\def\Qt{\tilde{Q}}
\def\Rt{\tilde{R}}
\def\Nt{\tilde{N}}

\def\at{\tilde{a}}
\def\st{\tilde{s}}
\def\ft{\tilde{f}}
\def\pt{\tilde{p}}
\def\qt{\tilde{q}}
\def\vt{\tilde{v}}
\def\nt{\tilde{n}}

\def\delb{\bar{\partial}}
\def\bz{\bar{z}}
\def\bD{\bar{D}}
\def\bB{\bar{B}}

\def\bk{{\bf k}}
\def\bl{{\bf l}}
\def\bp{{\bf p}}
\def\bq{{\bf q}}
\def\br{{\bf r}}
\def\bx{{\bf x}}
\def\by{{\bf y}}
\def\bR{{\bf R}}
\def\bV{{\bf V}}

\def\d{\delta}\def\D{\Delta}\def\ddt{\dot\delta}

\def\pa{\partial} \def\del{\partial}
\def\xx{\times}
\def\uno{\mbox{1 \kern-.59em {\rm l}}}
\def\RE{\mbox{R \kern-1.19em {\rm I}}}
\def\IM{\mbox{I \kern-.79em {\rm I}}}
\def\betaR{\beta_{\sst R}}
\def\hf{{\sst \frac{1}{2}}}

\def\trp{^{\top}}
\def\inv{^{-1}}
\def\dag{{^{\dagger}}}
\def\pr{^{\prime}}

\def\rar{\rightarrow}
\def\lar{\leftarrow}
\def\lrar{\leftrightarrow}

\newcommand{\0}{\,\!}      %this is just NOTHING!
\def\one{1\!\!1\,\,}
\def\im{\imath}
\def\jm{\jmath}

\newcommand{\tr}{\mbox{tr}}
\newcommand{\slsh}[1]{/ \!\!\!\! #1}

\def\vac{|0\rangle}
\def\lvac{\langle 0|}

\def\hlf{\frac{1}{2}}
\def\ove#1{\frac{1}{#1}}

\def\Box{\square}
\def\ZZ{\mathbb{Z}}
\def\CC#1{({\bf #1})}
\def\bcomment#1{}
\def\bfhat#1{{\bf \hat{#1}}}
\def\VEV#1{\left\langle #1\right\rangle}

\newcommand{\ex}[1]{{\rm e}^{#1}} \def\ii{{\rm i}}

\def\rr{{\rm r}} \def\rs{{\rm s}}\def\rv{{\rm v}}
\def\ri{{\rm i}}\def\rj{{\rm j}}

\newcommand{\lrbrk}[1]{\left(#1\right)}
\newcommand{\sfrac}[2]{{\textstyle\frac{#1}{#2}}}

\font\mybb=msbm10 at 12pt
\def\bb#1{\hbox{\mybb#1}}

\font\myBB=msbm10 at 18pt
\def\BB#1{\hbox{\myBB#1}}

\begin{document}

\hfill{ hep-th/0512194}\\
\rightline{IPPP/06/07}\\
\rightline{DCPT/06/14}

\vspace{20pt}

\begin{center}

{\Large \bf Amplitudes in the $\beta$-deformed Conformal Yang-Mills }

\vspace{30pt}

{\bf Valentin V. Khoze }

{\small \em
\begin{itemize}
\item[\ \ \ \ \ \ ] Department of Physics,
        University of Durham, Durham, DH1 3LE, UK
\end{itemize}
}

\vspace{10pt}

{\sffamily \tt
        valya.khoze@durham.ac.uk }

\vspace{30pt}
{\bf Abstract}
\end{center}

{\noindent
We study perturbative amplitudes in a large class
of theories obtained by marginal deformations of the $\cN=4$ supersymmetric Yang-Mills.
We find that planar amplitudes in the deformed theories are closely related to planar
amplitudes in the original $\cN=4$ SYM. For some classes of deformations the amplitudes
essentially coincide with the $\cN=4$ amplitudes to all orders in planar perturbation theory.
For more general classes of marginal deformations, the equivalence holds at up to four loops,
and at five loops it is likely to break down.
This implies that the iterative structure of planar MHV amplitudes recently discovered
by Bern, Dixon and Smirnov in \cite {BDS} for the $\cN=4$ theory also manifests itself
in a wider class of theories.}

\vspace{0.5cm}

\setcounter{page}{0}
\thispagestyle{empty}
\newpage

%%%%%%%%%%%%%% ordinary document (end)%%%%%%%%%%%%%%%%%%%%%
\baselineskip=6mm

\section{Introduction} \label{sec:one}

The last two years have seen a remarkable progress in developing new conceptual
and calculational approaches to compute perturbative on-shell scattering amplitudes
in gauge theory. Witten has proposed a duality between the
maximally supersymmetric $\cN=4$ Yang-Mills theory and a twistor string theory
\cite{Witten}.
This has stimulated wide-ranging progress in investigating scattering amplitudes in gauge theory
at tree level and one-loop level, ranging from QCD to the $\cN=4$ supersymmetric Yang-Mills,
and using novel manifestly on-shell approaches -- the MHV rules $+$ generalized unitarity
\cite{MHVtree,MHVloop},
and the on-shell recursion relations
\cite{RECtree,RECloop}.
For a summary of recent results one
can consult on-line talks in \cite{QMconf}.

One of the exciting recent developments is the remarkable
proposal of Bern, Dixon and Smirnov \cite{BDS} which applies to
amplitudes in the $\cN=4$ superconformal Yang-Mills theory.
This proposal results in the formula which determines the structure of all $n$-point planar MHV amplitudes
to all orders in perturbation theory. It follows that planar MHV amplitudes in the
superconformal $\cN=4$ Yang-Mills have an iterative structure, such that the
kinematic dependence of all higher-loop
MHV amplitudes can be determined from the known one-loop results.
The main motivation of the present paper is to study how this result
can be carried over to a wider family of gauge theories obtained by deformations
of the $\cN=4$ Yang-Mills. We will be mostly interested in the marginal
deformations of the $\cN=4$ studied earlier in \cite{LS,marginal,LM} and
we will also mention some known results \cite{BKV} about orbifolds and orientifolds.

The proposal made in \cite{BDS} gives an exponentiated ansatz which predicts the
finite part of $n$-point planar MHV amplitudes in
the superconformal $\cN=4$ Yang-Mills to all orders in planar perturbation theory.
The form of this ansatz is largely dictated
by explicit three-loop calculations of four-point amplitudes carried out in \cite{BDS},
and is also based on earlier work \cite{BDDK,BRY,ABDK,Smirnov}.
Importantly, it is also consistent with known results about exponentiation of IR divergencies
\cite{STY,MS,Catani} in a generic massless gauge theory.

$\cN=4$ SYM in the conformal phase is in many respects a special theory:
as a field theory it is maximally
supersymmetric, conformal and $SL(2,Z)$ self-dual. It also has a dual string theory
description \cite{AdSCFT,AdSCFT2}
provided by the AdS/CFT correspondence. From the perspective of the on-shell amplitudes
it is not clear at present which of these features (if any) are responsible
for the iterative structure uncovered in
\cite{BDS}. This makes it particularly interesting to investigate the
amplitudes in theories related to the $\cN=4$ SYM but with lower degree of symmetry.

There are at least two distinct ways of deforming the $\cN=4$ superconformal
Yang-Mills theory to obtain another gauge theory with a lower degree of
(super)-symmetry.
One class of interesting deformations is the orbifold projection \cite{BKV} of
the $\cN=4$ SYM.
Orbifolding involves imposing a discrete projection operator on the parent
$\cN=4$ theory.
This procedure modifies the field content of the theory:
only the fields which are invariant under a discrete symmetry subgroup of the
original theory are retained and other fields
are projected out. Deformations of the $\cN=4$ SYM by orbifolding reduces the
amount of supersymmetry in the daughter theory to $\cN=2$, $\cN=1$ or $\cN=0$.
Remarkably, it was shown in \cite{BKV} that
 in the large $N$ limit the parent and the daughter theories are
perturbatively equivalent\footnote{Non-perturbative large $N$ equivalence
of orbifolded theories is still
a matter of debate \cite{Strassler,ASV}.}.
This implies that planar perturbative amplitudes in the orbifold projections of
$\cN=4$ have exactly the same iterative structure as the one uncovered in \cite{BDS}
for $\cN=4$.

Another method is to deform the $\cN=4$ SYM by orientifolding \cite{ASV}.
The resulting orientifold daughter theory is a non-supersymmetric Yang-Mills
with the
$SU(N)$ gauge field plus 6 adjoint scalars and 4 Dirac fermions transforming in the (anti)symmetric
representation.
This theory is conformal at large $N$ and the planar equivalence with the $\cN=4$ SYM
is expected to hold in this case also non-perturbatively.\footnote{I thank Adi Armoni and Lance Dixon
for useful comments on orbi- and orientifolding of the $\cN=4$.}

The second (and more non-trivial) way of modifying $\cN=4$ is to deform it
by adding marginal deformations \cite{LS,marginal} to the Lagrangian.
This leads to a conformally invariant theory with a lower $\cN=1$ supersymmetry.
The marginal deformations provide a much richer structure of deformed theories compared to
orbifold projections. There is now a continuous family
of deformations and the original $\cN=4$ SYM corresponds a point on the moduli space
of the deformed theories \cite{LS}. In this paper we will investigate relations between planar
perturbative amplitudes
in the $\cN=4$ and in its marginal deformations. For related perturbative calculations
of off-shell correlators in $\beta$-deformed theories see \cite{FG,PSZ,MPSZ}.

The paper is organized as follows.
In Section {\bf 2} we collect some known facts about the $\cN=4$ supersymmetric Yang-Mills
and its marginal $\beta$-deformations.
In Section {\bf 3} we discuss a special class of such deformations, which is characterized
by a {\it real} deformation parameter $\beta=\betaR.$
In the large $N$ limit these $\betaR$-deformations form a one-real-parameter family of $\cN=1$ conformal
gauge theories. We prove that all planar amplitudes in these theories are equal to the corresponding
amplitudes in the original $\cN=4$ theory (times an overall phase factor which we determine).
The proof holds to all orders in planar perturbation theory and applies to all amplitudes
(MHV and non-MHV and for any number of external legs). This implies that the iterative structure
of $\cN=4$ amplitudes \cite{BDS} is carried over to the family of theories deformed by $\betaR$.
Section {\bf 3.1} first illustrates how this proof works for some specific diagrams,
and in Section {\bf 3.2} we give a more abstract general proof of the planar
equivalence using star-products.

In Section {\bf 4} we consider more general marginal deformations with deformation
parameters being complex. We show that all planar amplitudes in the deformed theory
match those in the $\cN=4$ SYM up to and including the four-loop level. At five loops,
however, we find a mismatch which very likely will result
in a breakdown in the iterative structure of planar amplitudes in the
general deformed theory at five loops and beyond.
We present our conclusions in Section {\bf 5}.

\section{Conformal deformations of $\N = 4$ SYM} \label{sec:two}

Our starting point is the $\N =4$ supersymmetric gauge theory.
In terms of $\N=1$ superfields, the superpotential of the theory is
\be
i g \, \Tr ( \Phi_1 \Phi_2 \Phi_3 - \Phi_1 \Phi_3 \Phi_2)
\label{superpot1}
\ee
Here $\Phi$'s are three chiral
superfields transforming in the adjoint representation of the gauge group.
In components, the $\N =4$ Lagrangian reads:
\bea
{\cal{L}}={\cal{L}}_b+{\cal{L}}_f
\eea
where
\bea
\label{Lbose}
{\cal{L}}_b = \Tr \left( {1 \over 4}
F^{\mu \nu}F_{\mu \nu} +
(D^\mu \bar \Phi^i ) (D_\mu \Phi_i  )-{g^2\over 2}[\Phi_i,\Phi_j][\bar \Phi^i,\bar \Phi^j]
+{g^2\over 4}[\Phi_i,\bar \Phi^i][\Phi_j,\bar \Phi^j]\right)
\eea
and
\beqa
\nonumber
\label{Lfermi}
{\cal{L}}_f=\Tr \left(  \l_{A} \s^{\mu} D_{\mu} \bar\l^A
- i g([\l_4,\l_i]\bar \Phi^i+[\bar\l^4,\bar \l^i]\Phi_i)
+{i g\over 2}(\epsilon^{ijk}[\l_i,\l_j]\Phi_k
+\epsilon_{ijk}[\bar\l^i,\bar\l^j]\bar \Phi^k)\right)
\\
\eeqa
We use $A=1,\ldots , 4$ and $i,j,k=1, \ldots , 3$, such that
$\Phi_{i}$ are the three complex scalar field components of the chiral superfields $\Phi_i$,
and $\lambda_{i}$ are their fermionic superpartners. The fourth fermion $\lambda_{4}$ is the
superpartner of the gluon.
The $SU(N)$ generators are normalised as
$\Tr \left( T^a T^b \right) = \delta^{ab}.$

We now turn to discuss marginal deformations of this theory.
There is a 3-complex-parameter family of such deformations which preserve $\N=1$
supersymmetry \cite{LS}.
% but not necessarily the conformal invariance of the original $\N=4$ SYM.
These deformations are described  \cite{marginal}, \cite{LS}
by replacing \eqref{superpot1} with the superpotential
\be \label{superpotgen}
i h \, \Tr( e^{ i  \pi \beta} \Phi_1\Phi_2 \Phi_3 - e^{- i  \pi \beta}
 \Phi_1\Phi_3 \Phi_2  ) + ih'\, \Tr( \Phi_1^3 +\Phi_2^3 +\Phi_3^3 )
\ee
At the classical level the deformations are marginal (since all operators in
the component Lagrangian have classical mass dimension four)
and the deformed theory is parameterized by four complex constants,
$h, h' , \beta, \tau,$
with $\tau$ being the usual complexified gauge coupling. At the quantum level,
this deformation is not exactly marginal since the operators
in \eqref{superpotgen} can develop anomalous dimensions. Using constraints of
$\N=1$ supersymmetry,  and the exact
NSVZ beta function approach \cite{NSVZ},
Leigh and Strassler \cite{LS} made an important observation. They
showed that the deformation \eqref{superpotgen}
is marginal at the quantum level
subject to a {\it single} complex constraint on the four parameters,
$\gamma(h, h' , \beta, \tau)=0.$ Here the function $\gamma$ is the sum of the
anomalous dimensions $\gamma_i$ of the three fields $\Phi_i$, so that
$\gamma= \sum_{i=1,2,3} \gamma_i.$
This constraint implies that there is a 3-complex-dimensional surface
$\gamma(h, h' , \beta, \tau)=0$
of conformally invariant $\N=1$ theories obtained by deforming $\N=4$ SYM.

In this paper we set $h'=0.$ In this case, the deformed theory
preserves a global $U(1)\times U(1)$ symmetry,
\bea \nonumber
U(1)_1:& ~~~~&(\Phi_1,\Phi_2 ,\Phi_3 ) \to (\Phi_1,e^{i\varphi_1}\Phi_2 ,e^{-i \varphi_1} \Phi_3 )
\\ \label{u1s}
U(1)_2:&~~~~&(\Phi_1,\Phi_2 ,\Phi_3 ) \to (e^{-i \varphi_2} \Phi_1,e^{i\varphi_2}\Phi_2 , \Phi_3 )
\eea
where $\Phi_i$ are chiral $\N=1$ superfields.

At present, the complete solution of the Leigh-Strassler constraint,
\be
\gamma(h,  \beta, \tau)= \, 0
\label{LSconstr}
\ee
is unknown, i.e. the function $\gamma= \sum_{i=1}^{3} \gamma_i(h, h' , \beta, \tau)$
in \eqref{LSconstr} is unknown beyond the
two-loop order in perturbation theory \cite{marginal}. One approach, followed in \cite{KZ},
is to work to the leading order in perturbation theory in deformations, in which case
$\gamma$ is known.
This motivated the authors of \cite{KZ} to expand the
$e^{ \pm i  \pi \beta}$ factors in \eqref{superpotgen}
and to work to the first order in all deformations.
We will follow a slightly different approach which does not require to expand the $\beta$-dependent factors.
We will discuss a fixed-order in $h$ and $g$ perturbative treatment of general conformal
deformations in Section {\bf \ref{sec:four}}.

But first we consider in Section {\bf \ref{sec:three}} a restricted class of deformations
which will allow us to work to all-orders in perturbation theory.

\section{Planar amplitudes in real-$\beta$ deformations} \label{sec:three}

Here we restrict the general deformation \eqref{superpotgen} to the
`phase-deformation' previously considered also by Lunin and Maldacena in \cite{LM}.
We set $h=g,$ $h'=0$ and keep $\beta$ real, $\beta=\betaR \in {\bf R}$.
The deformation of the
superpotential of the $\N=4$ theory then reads
\be
\label{superpot2}
i g \,\Tr( \Phi_1 \Phi_2 \Phi_3 -\Phi_1 \Phi_3 \Phi_2 ) \to
i g \,\Tr( e^{ i \pi \betaR } \Phi_1 \Phi_2 \Phi_3 - e^{-i \pi \betaR } \Phi_1 \Phi_3 \Phi_2 )\ .
\ee
This superpotential preserves the $\N=1$ supersymmetry of the original $\N=4$ SYM
and leads to a theory with a global $U(1)\times U(1)$ symmetry \eqref{u1s}.
Less obvious is the fact (first proven in Ref.~\cite{MPSZ})
that \eqref{superpot2} describes an exactly marginal deformation
of the theory in the limit of large number of colors.
As a by-product of our general discussion of planar amplitudes,
in Section {\bf 3.2}
we will reproduce this result of \cite{MPSZ} and show that the
planar perturbation theory of the deformed model is conformal, and the corresponding
Leigh-Strassler constraint is satisfied at leading order in the $N \to \infty$ limit.

The main goal of this Section is to analyze scattering amplitudes of the theory
deformed via \eqref{superpot2} to all orders in planar perturbation theory,
(and to all orders in $\betaR$)
and specifically compare them to the corresponding amplitudes in the original
$\N=4$ SYM.

Lunin and Maldacena pointed out \cite{LM} that
the deformation of the superpotential \eqref{superpot2}
can be viewed as arising from a new definition of the products between all fields in the
$\N=4$ Lagrangian,
\be
\label{star}
f * g \equiv
e^{ i  \pi \betaR (Q_1^f Q_2^g - Q_2^f Q_1^g) } f g
\ee
where $fg$ is an
ordinary product
and $(Q_1^{\rm field},Q_2^{\rm field})$ are the $U(1)_1 \times U(1)_2$
charges of the fields ($f$ or $g$).
The values of the charges for all fields are read from \eqref{u1s}:
\bea
\Phi_1 \,(\Phi_1, \lambda_1)\,  : \,\,\,\, &&(Q_1\, , Q_2)=(0\, , -1)
\label{ch1}\\
\Phi_2 \, (\Phi_2, \lambda_2)\,  : \,\,\,\, &&(Q_1\, , Q_2)=(1\, , 1)
\label{ch2}\\
\Phi_3 \, (\Phi_3, \lambda_3)\,  : \,\,\,\, &&(Q_1\, , Q_2)=(-1\, , 0)
\label{ch3}\\
 V \, (A_\mu, \lambda_4)\,  : \,\,\,\, &&(Q_1\, , Q_2)=(0\, , 0)
 \label{ch4}
\eea
and for the conjugate fields ($\bar\Phi_i$ and $\bar\lambda_i$) the charges are opposite.

For example, to calculate the star-product of three fields, $\Phi_1 * \Phi_2 * \Phi_3$,
we first find $\Phi_2 * \Phi_3= e^{ i  \pi \betaR (0+1)} \Phi_2 \Phi_3$ using
\eqref{ch1},\eqref{ch2},\eqref{star}.
Then $\Phi_1 * \Phi_2 * \Phi_3 = \Phi_1 * (\Phi_2 \Phi_3)e^{ i  \pi \betaR } =
\Phi_1\Phi_2 \Phi_3 e^{ i  \pi \betaR }$
since the charges of the composite field  $\Phi_2 \Phi_3$ are
$(Q_1, Q_2)=(0, 1),$  and as such cannot
contribute to the last product. This implies
\be
\Tr( \Phi_1 *\Phi_2* \Phi_3 -\Phi_1 *\Phi_3* \Phi_2 ) =\,
\Tr( e^{ i \pi \betaR } \Phi_1 \Phi_2 \Phi_3 - e^{-i \pi \betaR } \Phi_1 \Phi_3 \Phi_2 )
\ee
and confirms
that the deformation of the superpotential
Eq.~\eqref{superpot2} is induced by the star-product \eqref{star}.

With the star-product \eqref{star}, the component Lagrangian of the
$\betaR$-deformed theory \eqref{superpot2} immediately follows from
\eqref{Lbose},\eqref{Lfermi}. Only the third term in \eqref{Lbose} and the
last two terms in \eqref{Lfermi} change, such that in total we have
\bea
\nonumber
&&{\cal{L}} = \, \Tr \Bigg( {1 \over 4}
F^{\mu \nu}F_{\mu \nu} +
(D^\mu \bar \Phi^i ) (D_\mu \Phi_i  )
- {g^2\over 2} [\Phi_i,\Phi_j]_{\beta_{ij}}[\bar \Phi^i,\bar \Phi^j]_{\beta_{ij}}
+{g^2\over 4}[\Phi_i,\bar \Phi^i][\Phi_j,\bar \Phi^j] \\
&&+ \l_{A} \s^{\mu} D_{\mu} \bar\l^A
- i g ([\l_4,\l_i]\bar \Phi^i+[\bar\l^4,\bar \l^i]\Phi_i)
+{i g\over 2}(\epsilon^{ijk}[\l_i,\l_j]_{\beta_{ij}}\Phi_k
+\epsilon_{ijk}[\bar\l^i,\bar\l^j]_{\beta_{ij}}\bar \Phi^k)\Bigg)
\nonumber \\
\label{Ldef}
\eea
Here we introduced the $\betaR$-deformed commutator of fields which is simply
\be
[ f_i,g_j]_{\beta_{ij}} :=\,
e^{ i \pi \beta_{ij} }\, f_i g_j -\,
e^{ -i \pi \beta_{ij} }\, g_j f_i \ . \label{combeta}
\ee
and $\beta_{ij}$ is defined as
\be
\beta_{ij}=-\beta_{ji} \, , \quad
\beta_{12}=\, -\beta_{13}=\, \beta_{23} :=\, \betaR \ .
\label{betaijs}
\ee
It is now easy to see which of the Feynman vertices are affected by $\betaR$.
All $\betaR$-dependent four-scalar interactions are contained in the term
\be
\label{phi4beta}
[\Phi_i,\Phi_j]_{\beta_{ij}}[\bar \Phi^i,\bar \Phi^j]_{\beta_{ij}} = \,
e^{ i 2\pi \beta_{ij} }\, \Phi_i\Phi_j \bar \Phi^i \bar \Phi^j + \,
e^{ -i 2\pi \beta_{ij} }\, \Phi_j\Phi_i \bar \Phi^j \bar \Phi^i - \,
\Phi_i\Phi_j \bar \Phi^j \bar \Phi^i -\, \Phi_j\Phi_i \bar \Phi^i \bar \Phi^j
\ee
The first and second terms on the right hand side give the color-ordered Feynman
vertex depicted in the first diagram in Figure 1.
\begin{figure}[t]
\label{fig:1beta}
\psfrag{r}{\LARGE${\bar\Phi_j}$}
\psfrag{j}{\LARGE${\bar\lambda_j}$}
\psfrag{h}{\LARGE${\lambda_i}$}
\psfrag{g}{\LARGE${\lambda_j}$}
\psfrag{k}{\LARGE${\bar\lambda_i}$}
\psfrag{A}{\LARGE${- g^2  \times e^{ 2\pi i \beta_{ij} }}$}
\psfrag{B}{\LARGE${i  g\, \epsilon^{ijk}} \times e^{ \pi i \beta_{ij} }$}
\psfrag{C}{\LARGE${i g\, \epsilon^{ijk}}\times e^{ \pi i \beta_{ij} }$}
\psfrag{a}{\LARGE${\Phi_i}$}
\psfrag{b}{\LARGE${\Phi_j}$}
\psfrag{f}{\LARGE${\bar\Phi_i}$}
\psfrag{c}{\LARGE${\Phi_k}$}
\psfrag{d}{\LARGE${\bar\Phi_k}$}
\begin{center}
\scalebox{0.55}{\includegraphics{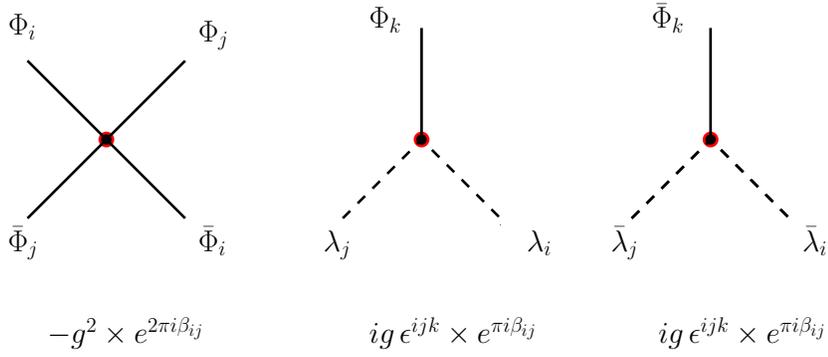}}
\end{center}
\caption{\it All $\betaR$-dependent Feynman vertices in color-ordered
perturbation theory.
}
\end{figure}
\begin{figure}[h]
\label{fig:2beta}
\psfrag{A}{\LARGE${{1 \over 2}\,g^2 }$}
\psfrag{B}{\LARGE${{1 \over 4}\, g^2}$}
\psfrag{C}{\LARGE${{1 \over 4}\, g^2}$}
\psfrag{a}{\LARGE${\Phi_i}$}
\psfrag{b}{\LARGE${\Phi_j}$}
\psfrag{f}{\LARGE${\bar\Phi_j}$}
\psfrag{r}{\LARGE${\bar\Phi_i}$}
\begin{center}
\scalebox{0.55}{\includegraphics{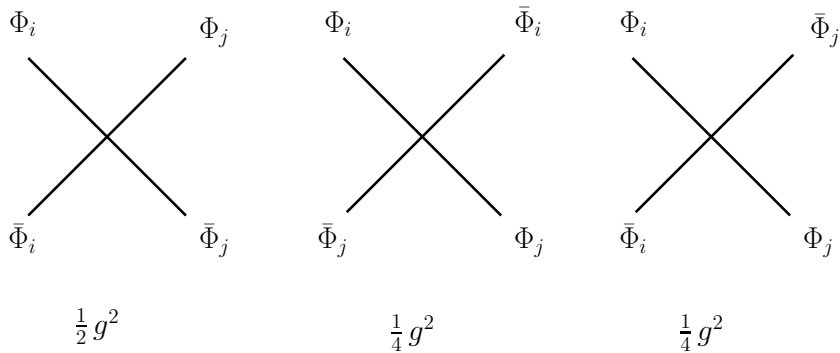}}
\end{center}
\caption{\it $\betaR$-independent $\phi^4$ color-ordered vertices.
}
\end{figure}
The $\betaR$-independent third and fourth terms in \eqref{phi4beta} contribute to the
first diagram in Figure 2. The remaining four-scalar interactions in
\eqref{Ldef} are determined by
$(1/2)[\Phi_i,\bar \Phi^i][\Phi_j,\bar \Phi^j],$ and do not depend on $\beta.$
All the $\betaR$-independent $\phi^4$ interactions are given by the three color-ordered
vertices in Figure 2.

The only remaining $\betaR$-dependent interactions are the Yukawa interactions --
the last terms
on the right hand side of Eq.~\eqref{Ldef}. They give color-ordered Feynman vertices
depicted in the second and third diagrams in Figure 1.

\subsection{Planar amplitudes} \label{sec:planar}

Employing Feynman rules derived above we can make the following statements about
scattering amplitudes in the $\betaR$-deformed theory.

\begin{enumerate}

\item Any Feynman vertex involving a gluon, $g,$ or a gluino, $\bar\lambda_4,$ $\lambda_4,$
is unmodified. These vertices plus the three scalar interactions in Figure 2
give the complete
set of the $\betaR$-independent vertices.

\item Three vertices in Figure 1 are the only vertices which depend on the
deformation parameter $\betaR$. For a Feynman diagram in the $\betaR$-deformed theory
to agree with the corresponding diagram in the $\N=4$ SYM, it should either not contain
these three types of vertices, or the phases introduced by these vertices should
mutually cancel.
Note that the phase depends on the order of the fields in the color-ordered vertices,
for example the vertex of the type
$\lambda_1 \lambda_2 \Phi_3 \sim e^{i \pi \beta_{12}}=e^{i \pi \betaR} $
while the vertex
$\lambda_2 \lambda_1 \Phi_3 \sim e^{i \pi \beta_{21}}=e^{-i \pi \betaR}$.

\item All tree-level amplitudes in the deformed theory coincide with the corresponding
amplitudes in the $\N=4$ theory up to a multiplicative constant factor.
This constant depends only on the particle types in the external lines, and not on the
internal structure of the diagram.

\item Tree-level and one-loop
amplitudes with only gluons on the external lines are trivially identical to
those in $\N=4$. This also applies to amplitudes with external $g$'s, $\bar\lambda_4$'s, and
$\lambda_4$'s.

\item Beyond one-loop, non-planar diagrams of the $\betaR$-deformed theory in general
can contain non-trivial phase factors which depend on the structure of internal
lines of individual diagrams.

\item A key point which we will prove below is
that the total phase factor associated with any given planar amplitude
is entirely determined by the external lines and does not depend on topologies and types of
internal (loop) interactions.
In other words, any planar loop amplitude in the $\betaR$-deformed theory is equal to the
corresponding amplitude in the original $\N=4$ SYM times an overall external phase factor.
In particular, this universal phase factor can be easily determined using star-products
of external fields (as will be explained in Section {\bf \ref{sec:starproof}}), or it can also be read
off the corresponding tree-level amplitude, if it is non-vanishing.

\item It also follows that  MHV amplitudes to all orders in planar perturbation theory are proportional to
the corresponding tree-level amplitudes. The $\betaR$-dependence of these
planar all-orders amplitudes is entirely contained in the tree-level amplitude
prefactor.

\end{enumerate}

We start by considering planar
multi-loop amplitudes with only gluons and gluinos: $\bar\lambda_4$ and
$\lambda_4$ on the external lines. From the comments above it follows that
these amplitudes in the deformed theory should be the same as in the original $\N=4$.

To verify this statement we
first consider such diagrams where the $\betaR$-phases are carried only by the 4-scalar
interactions -- the first vertex in Figure 1. What distinguishes this
vertex from the rest of 4-scalar vertices (of Figure 2) is that it is of
the alternating $i-j-i-j$ type (compared to the consecutive $i-i-j-j$ structure
of three vertices in Figure 2). This index structure of the four scalar vertices
is illustrated in Figure 3. It follows that only the $\beta$-dependent first
vertex in Figure 3 corresponds to intersecting index-lines. In this sense the
$\betaR$-dependent vertex is `non-planar'.

\begin{figure}[h]
\label{fig3beta}
\psfrag{i}{\LARGE${i}$}
\psfrag{j}{\LARGE${j}$}
\begin{center}
\scalebox{0.55}{\includegraphics{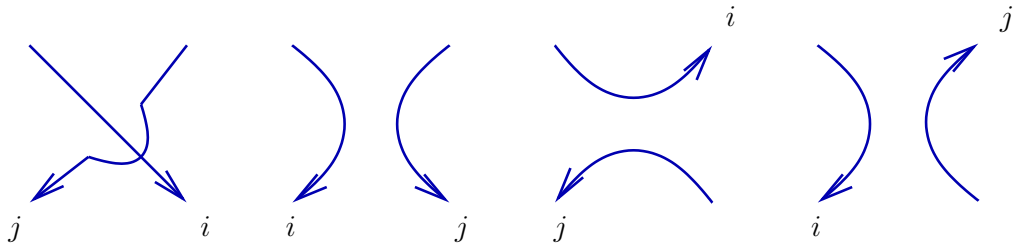}}
\end{center}
\caption{\it Index-lines representation of the four scalar vertices.
The first diagram represents the $\betaR$-dependent first vertex in Figure 1.
The remaining three diagrams correspond to the three vertices in Figure 2.
}
\end{figure}

Since gluons cannot join together scalar fields of two different flavors,
$i$ and $j,$ the first vertex can appear either an even number of times, or
it must be accompanied by a non-planar crossing. In planar diagrams,
we can form closed planar chains of $\phi^4$ interactions and discover
that all the individual phases cancel each other
leaving us with the $\N=4$ result.

Now we add diagrams with the $\betaR$-dependent Yukawa interactions.
It can be shown
that all individual phases will precisely cancel each other for planar amplitudes
with external gluons,
leaving behind the $\N=4$ expression. Instead of verifying this result on a diagram-by-diagram basis,
in the following section {\bf \ref{sec:starproof}} we will prove it in general using
properties of star-products. But it may be instructive to first illustrate this phase cancellation
with a simple example.
Consider
a 2-loop planar diagram with external gluons depicted in Figure 4.
\begin{figure}[t]
\label{figdbox}
\psfrag{a}{\LARGE${g}$}
\psfrag{A}{\LARGE${\bar\Phi_3}$}
\psfrag{B}{\LARGE${\Phi_3}$}
\psfrag{C}{\LARGE${\lambda_1}$}
\psfrag{D}{\LARGE${\bar\lambda_1}$}
\psfrag{E}{\LARGE${\lambda_2}$}
\psfrag{F}{\LARGE${\bar\lambda_2}$}
\begin{center}
\scalebox{0.45}{\includegraphics{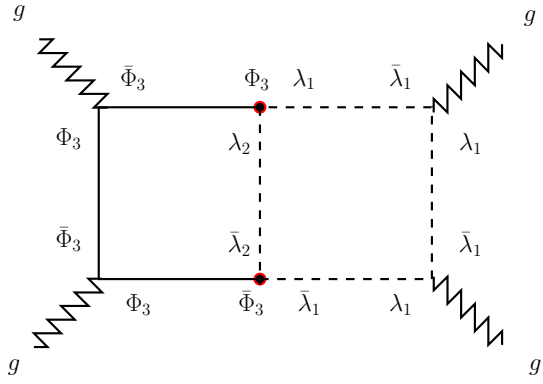}}
\end{center}
\caption{\it Example of a 2-loop diagram contributing to the four-gluon amplitude.
Red dots denote two $betaR$-dependent Yukawa vertices. Their phases
cancel and the total contribution is $\betaR$-independent.
}
\end{figure}
There are two deformed Yukawa vertices in this diagram. The top vertex is
$\lambda_1 \lambda_2 \Phi_3 \sim g e^{i\pi \beta_{12}} \sim  g e^{i\pi \betaR},$
and the bottom vertex is
$\bar\lambda_2 \bar\lambda_1 \bar\Phi_3 \sim g e^{i\pi \beta_{21}} \sim  g e^{-i\pi \betaR}.$
The two phase factors cancel each other and the diagram is $\betaR$-independent,
as advertised.
Discussion of general diagrams is deferred to Section {\bf \ref{sec:starproof}}.

When external lines include fields from $\Phi_i$ supermultiplets,
the phase-dependence will remain. Nevertheless, at planar level, this
dependence can be completely fixed
by the external lines configuration in the following way.
For every n-point planar color-ordered amplitude $A_n$ write down a factor
of $\Tr(F_1 *F_2 * \ldots *F_3)$ where $F_i$ denote all external fields in $A_n$
\be
\label{starphase}
A_n(F_1,F_2, \ldots, F_n)\,: \qquad \Tr(F_1 *F_2 * \ldots *F_n) = [{\rm phase}(\betaR)] \,\times\,
\Tr(F_1 F_2  \ldots F_n)
\ee
Then the $\betaR$-phase of the amplitude is equal to the [phase$(\betaR)$]
of this star-product (see Section {\bf \ref{sec:starproof}}).

To illustrate this, we consider first two simple MHV planar amplitudes with four scalars,
$A_4(\Phi_1,\Phi_2,\bar\Phi_1,\bar\Phi_2)$ and $A_4(\Phi_1,\Phi_2,\bar\Phi_2,\bar\Phi_1).$
The $\betaR$-dependence of the first amplitude is determined via:
\be
\label{starph1st}
\Tr(\Phi_1*\Phi_2*\bar\Phi_1*\bar\Phi_2) =\, e^{2i \pi \betaR}\, \Tr(\Phi_1\Phi_2\bar\Phi_1\bar\Phi_2)
\ee
and of the second, via:
\be
\label{starph2nd}
\Tr(\Phi_1*\Phi_2*\bar\Phi_2*\bar\Phi_1) =\, 1\, \Tr(\Phi_1\Phi_2\bar\Phi_2\bar\Phi_1)
\ee
Examples of diagrams contributing to these amplitudes are shown in Figure 5.
\begin{figure}
\label{figbox}
\psfrag{a}{\LARGE${\Phi_1}$}
\psfrag{b}{\LARGE${\Phi_2}$}
\psfrag{c}{\LARGE${\bar\Phi_1}$}
\psfrag{d}{\LARGE${\bar\Phi_2}$}
\psfrag{A}{\LARGE${\lambda_2}$}
\psfrag{B}{\LARGE${\bar\lambda_2}$}
\psfrag{C}{\LARGE${\bar\lambda_4}$}
\psfrag{D}{\LARGE${\lambda_4}$}
\psfrag{E}{\LARGE${\lambda_1}$}
\psfrag{F}{\LARGE${\bar\lambda_1}$}
\psfrag{G}{\LARGE${\bar\lambda_3}$}
\psfrag{H}{\LARGE${\lambda_3}$}
\begin{center}
\scalebox{0.45}{\includegraphics{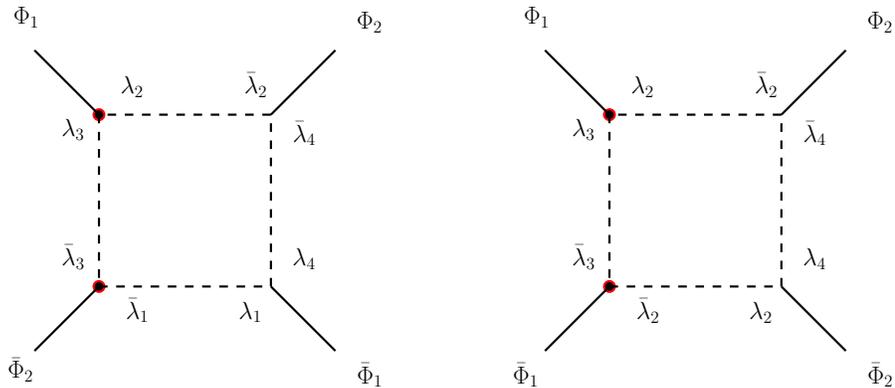}}
\end{center}
\caption{\it Two examples of 1-loop contributions to MHV amplitudes with four external scalars.
The amplitude on the left is $\beta$-dependent,
$A_4(\Phi_1,\Phi_2,\bar\Phi_1,\bar\Phi_2) \sim e^{2i \pi \betaR}$. In the second amplitude,
$A_4(\Phi_1,\Phi_2,\bar\Phi_2,\bar\Phi_1)\sim 1,$ the phases cancel.}
\end{figure}
It is easy to see that both $\betaR$-dependent Yukawa vertices contribute $e^{i \pi \betaR}$ each
for the first diagram, and $e^{i \pi \betaR}$ times $e^{-i \pi \betaR}$ for the second one.
Hence, $A_4(\Phi_1,\Phi_2,\bar\Phi_1,\bar\Phi_2) \sim e^{2i \pi \betaR}$ and
$A_4(\Phi_1,\Phi_2,\bar\Phi_2,\bar\Phi_1)\sim 1,$ in agreement with Eqs.~\eqref{starph1st},\eqref{starph2nd}.

We can further consider Next-to-MHV amplitudes with six scalars, depicted in Figure 6.
\begin{figure}
\label{fighexagon}
\psfrag{a}{\LARGE${\Phi_1}$}
\psfrag{b}{\LARGE${\bar\Phi_2}$}
\psfrag{c}{\LARGE${\Phi_3}$}
\psfrag{d}{\LARGE${\bar\Phi_1}$}
\psfrag{e}{\LARGE${\Phi_2}$}
\psfrag{f}{\LARGE${\bar\Phi_3}$}
\psfrag{A}{\LARGE${\lambda_3}$}
\psfrag{B}{\LARGE${\bar\lambda_3}$}
\psfrag{D}{\LARGE${\lambda_1}$}
\psfrag{C}{\LARGE${\bar\lambda_1}$}
\psfrag{E}{\LARGE${\lambda_2}$}
\psfrag{F}{\LARGE${\bar\lambda_2}$}
\psfrag{W}{\LARGE${\lambda_4}$}
\psfrag{Z}{\LARGE${\bar\lambda_4}$}
\begin{center}
\scalebox{0.45}{\includegraphics{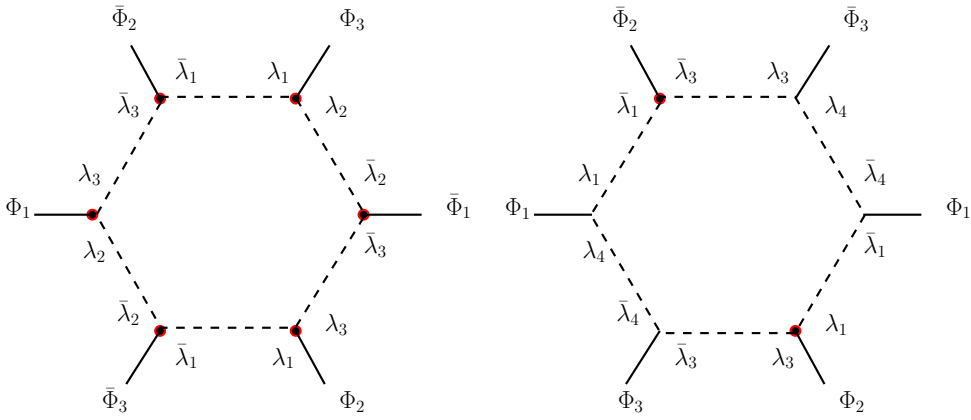}}
\end{center}
\caption{\it Two examples of Next-to-MHV amplitudes with six external scalars.
The amplitude on the left,
$A(\Phi_1,\bar\Phi_2,\Phi_3,\bar\Phi_1,\Phi_2,\bar\Phi_3,) \sim e^{-6i \pi \betaR}$,
and the amplitude on the right,
$A(\Phi_1,\bar\Phi_2,\bar\Phi_3,\Phi_1,\Phi_2,\Phi_3,) \sim e^{2i \pi \betaR}.$
}
\end{figure}
The phase-factors associated with each of these diagrams are again in
complete agreement with the general rule Eq.~\eqref{starphase}.

\subsection{All-orders proof based on star-products} \label{sec:starproof}

In this section we present a proof of the assertion
that the $\betaR$-dependence of a planar amplitude, $A_n(F_1, \ldots, F_n),$
is entirely determined by the configuration of external fields $F_1, \ldots F_n$
via Eq.~\eqref{starphase}.

We will rely on the fact the real-$\betaR$ deformation we are studying is equivalent
to the deformation by the star-product Eq.~\eqref{star}. The fact that the loop structure
of planar diagrams does not affect the total phase-factor induced by star-products,
is a well-known result in quantum field theories formulated on noncommutative spaces,
\cite{Filk,IIKK,Minwalla}.\footnote{This analogy with noncommutativity has already
prompted the authors of \cite{LM} to state that planar amplitudes in the $\betaR$-deformed theory
are same as in the original $\N=4$, see their Section 4.}
However, the noncommutative product \eqref{star} which
leads to the $\betaR$-deformed theory, is defined differently from the Moyal star-product
used in noncommutativity hence the proofs of planar equivalence presented
in \cite{Filk,IIKK,Minwalla} require a modification, which we now present.

Each color-ordered vertex in Feynman perturbation theory can be represented
by the corresponding term in the classical Lagrangian, e.g.
$\Tr(\Phi_j \Phi_i \bar\Phi_i \bar\Phi_j),$ or $\Tr(\lambda_i \lambda_j \Phi_k),$
or in general $\Tr(F_1 F_2  \ldots F_n)$ where $n$ is 3 or 4.
In the $\betaR$-deformed theory, this expression involves star-products,
$\Tr(F_1 *F_2 * \ldots *F_n).$ Any two vertices connected by a propagator at tree-level
can be represented by a single effective vertex, as shown in Figure 7. By this we mean
that the total phase factor of the connected tree diagram made out of these two vertices
is equal to the phase factor of the effective vertex.
Indeed, using the fact that each vertex $V$ in Figure 7 has a vanishing total
$Q$-charge and the associativity of the star-product, we have
\begin{figure}[t]
\label{figstar1}
\psfrag{a}{\LARGE${F_i}$}
\psfrag{b}{\LARGE${F_j}$}
\psfrag{c}{\LARGE${F_k}$}
\psfrag{d}{\LARGE${F_l}$}
\psfrag{F}{\LARGE${F_I}$}
\psfrag{B}{\LARGE${\bar{F}_I}$}
\begin{center}
\scalebox{0.55}{\includegraphics{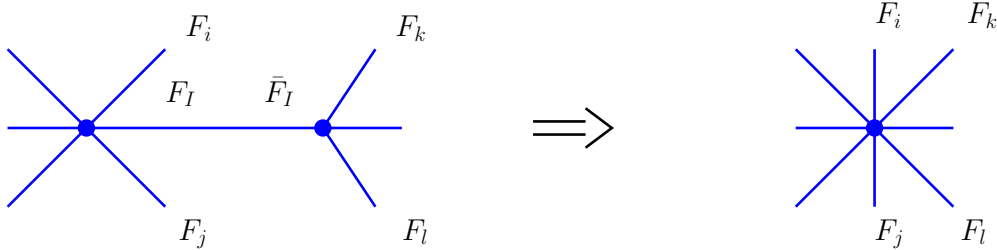}}
\end{center}
\caption{\it At tree level all vertices can be joined together into an effective
vertex without changing the total $\betaR$-phase of the diagram. $F$'s denote generic component
fields. We use the bar in $\bar{F}_I$ to emphasize that this field has opposite $Q$-charges
to the $F_I$ field on the other end of the internal propagator.
}
\end{figure}
\bea
&&V_{n} V_{m} \,= \,
(F_j*\ldots *F_i*F_I) \,(\bar{F}_I*F_k*\ldots*F_l) \\
&&=\, (F_j*\ldots *F_i*F_I) *(\bar{F}_I*F_k*\ldots*F_l)
\, =\, (F_j*\ldots *F_i)*(F_I *\bar{F}_I)*(F_k*\ldots*F_l) \nonumber
\eea
but since $F_I$ and $\bar{F}_I$ have opposite U(1) charges, we have $F_I *\bar{F}_I=\, 1 \,F_I \bar{F}_I \sim 1$
and this implies
\be
V_{n} V_{m} \sim
(F_j*\ldots *F_i*F_k*\ldots*F_l) = V_{n+m-2}
\ee
Hence an arbitrary tree amplitude has a phase factor which can be read of the
single effective vertex using the rule above. This verifies the statement in item (3) in the previous section.

Now we generalize to loops. Since a tree-level diagram can be reduced to an effective
vertex, a loop diagram can be represented by such a vertex with self-contractions
depicted in Figure 8.
A planar loop diagram must contain no intersecting lines.
\begin{figure}
\label{figstar2}
\begin{center}
\scalebox{0.55}{\includegraphics{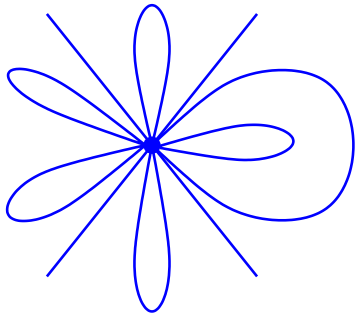}}
\end{center}
\caption{\it Reduced planar loop diagram is given by the corresponding effective tree-level
vertex with self-contraction.
Planarity implies that none of the (external and internal) lines can intersect.
Reduced diagram has the same $\betaR$-phase as the original Feynman diagram.
}
\end{figure}
It is easy to see that each planar self-contraction can be removed without affecting
the total $\betaR$-dependence of the diagram. This is because at the both ends of the
self-contraction we have fields of the opposite charges, and these fields are positioned
next to each other due to planarity, $F_I *\bar{F}_I \sim 1.$
Hence the phase-dependence of each planar diagram can be read off an effective tree-level vertex
in agreement with Eq.~\eqref{starphase}. This completes the proof.

We conclude that all planar amplitudes in the $\betaR$-deformed theory are proportional to their
$\N=4$ counterparts. Since, the $\N=4$ theory was finite, this implies the UV-finiteness
(or conformal invariance) of the deformed theory in the planar limit.

It follows that the intriguing iterative structure
of the planar MHV amplitudes in $\N=4$ superconformal Yang-Mills
uncovered in Ref.~\cite{BDS} is carried over to the phase-deformed theory
without modifications and to all orders in perturbation theory.
For generic MHV planar amplitudes,
the entire $\betaR$-dependence is contained in the tree-level
prefactor.

\section{More general conformal deformations} \label{sec:four}

We now consider more general deformations of $\N=4$ which do not require $\beta$ to be real.

\subsection{A non-supersymmetric generalization}

There is one simple class of extended deformations obtained by complexifying $\betaR$
in the component Lagrangian \eqref{Ldef}. The deformation parameter is now
a complex-valued $\beta$
\be
\label{betaC}
\beta=\betaR + i \chi \ , \qquad \betaR\ , \ \chi \in {\bf R}
\ee
The Lagrangian takes the same form as in \eqref{Ldef},
but the $\beta$-deformed commutators \eqref{combeta}
are written in terms of complex $\beta_{ij}$,
\be
\beta_{12}=\, -\beta_{13}=\, \beta_{23} :=\, \beta \in {\bf C} \ , \qquad
\beta_{ij}=-\beta_{ji}
\label{betaijs2}
\ee
We note that this theory (which can also be obtained by complexifying the star-product
of {\it component} fields) is non-supersymmetric and
is different from the $\N=1$ supersymmetric deformations\footnote{In the $\overline{\cal W},$ which
is the conjugate of the superpotential
${\cal W}$ in \eqref{superpotgen}, we would have to use $\beta^*$ which is the complex conjugate of $\beta$.
However the non-supersymmetric theory \eqref{Ldef} with complex $\beta$ depends only on
$\beta$ and not on $\beta^*.$} described by \eqref{superpotgen}
with complex $\beta.$ All the results of Section {\bf 3} including precise cancellation of
$\beta$ dependence in planar diagrams of gluons are carried over without modifications
to this complexified family of deformations. Hence we conclude that
there is a two-real-dimensional family of transformations of $\N=4$ which is nonsupersymmetric,
but has exactly the same (up to an overall complexified phase factor)
planar amplitudes as the $\N=4$ theory to all orders in perturbation theory.

This `planar equivalence' of perturbative amplitudes in the non-supersymmetric deformation
and in the $\N=4$ parent, however, should not be interpreted as the large-$N$ equivalence of the
two theories. The arguments of Ref.~\cite{DKR} imply that generic non-supersymmetric $\beta$-deformed
theories are not conformal and generate a scale due to couplings with double-trace operators.
This means that perturbative vacuum is not a true vacuum of the theory, and the `planar equivalence'
in the non-supersymmetric case should be viewed only as a formal relation between
perturbative amplitudes.

\subsection{$\cN=1$ supersymmetric $\beta$-deformations}

The second class of
more general deformations of $\N=4$ is described by the
superpotential
\be \label{superpotg2}
{\cal W}= \, i h \, \Tr( e^{ i  \pi \beta} \Phi_1\Phi_2 \Phi_3 - e^{- i  \pi \beta}
 \Phi_1\Phi_3 \Phi_2  )
\ee
Here $h$ and $\beta$ are complex parameters, and $\beta$ is given in \eqref{betaC}.
For a generic deformation \eqref{superpotg2}
which preserves conformal invariance of the theory only at the classical level,
it is pretty clear that the deformations will not cancel from planar diagrams --
they are not simple phases after all. But there may exist a condition on the parameters
$h$ and $\beta$ for which such cancellations do occur at least in the first few orders of
perturbation theory.
Can this condition be identical to the CFT constraint \eqref{LSconstr}?

The goal of this Section is to show that this is indeed the case, and that
in the deformed theory all planar amplitudes with external gluons are equal to those
in the $\N=4$ SYM up to a five-loop level. At five loops and beyond we expect
that planar amplitudes in the two theories begin to differ.

The UV finiteness (or equivalently the conformal invariance) of the theory deformed by
the superpotential
\be \label{spot3}
{\cal W} \,=\, \frac{1}{3!}\, C^{ijk}_{abc} \,\Phi^a_i \Phi^b_j \Phi^c_k
\ee
is guaranteed if the following condition \cite{marginal} is satisfied
\be \label{gamma1c}
C^{ijk}_{abc}\, C^{* \,d bc}_{\,\,\,ljk} \, =\, 2 g^2\, \delta^i_l \, \delta^d_a \, T_R
\ee
Here all $\Phi$'s transform in the adjoint representation and $T_R = N.$
The constraint \eqref{gamma1c} follows from the one-loop finiteness of the theory,
but it also automatically guarantees the UV finiteness of the
$\cN=1$ theory at two-loop level
\cite{marginal}. Hence, \eqref{gamma1c} is the solution of the Leigh-Strassler
CFT constraint \eqref{LSconstr} at
the two-loop level.

Our superpotential \eqref{superpotg2},
corresponds to
\be
C^{123}_{abc}\, \sim \, - \, h \left(f^{abc} (e^{i\pi\beta} + e^{-i\pi\beta})
+ i d^{abc} (e^{i\pi\beta} - e^{-i\pi\beta})\right)
\ee
and the two-loop CFT constraint \eqref{gamma1c} gives \cite{FG,PSZ}\footnote{In deriving
\eqref{con2lN} we used Eq.~\eqref{betaC} and the properties of the
$SU(N)$ structure constants, $f^{abc}f^{dbc}= \delta^{ad} N$ and
$d^{abc}d^{dbc}= \delta^{ad} (N-4/N).$}
\be
\label{con2lN}
|h|^2 \left( {\rm cosh}(2\pi \chi) +\frac{2}{N^2} {\rm cos}(2\pi \betaR) -
\frac{2}{N^2}{\rm cosh}(2\pi \chi) \right)=\, g^2
\ee
In the planar approximation we neglect $1/N^2$ terms and the constraint
simplifies,
\be
\label{con2l}
|h|^2 {\rm cosh}(2\pi \chi) =\, g^2
\ee
An important feature of the planar constraint \eqref{con2l}
is that it does not depend on
$\betaR$ and, furthermore, for real $\beta$ it collapses to $|h|^2=g^2.$
This
verifies at the two-loop level the fact that the real-$\betaR$-deformation
is an exactly marginal deformation in the planar limit -- in agreement with the
more powerful all-orders discussion
given earlier in  Section {\bf 3}. As a bonus, we also see that subleading in $1/N^2$
terms in \eqref{con2lN} (which do not affect planar diagrams) do in fact depend on $\betaR$.

We are now ready to concentrate on the
amplitudes with external gluons in planar perturbation theory, which
is the main subject of this Section.
These amplitudes can be organized in a simple way using
supergraph Feynman perturbation theory. The relevant vertices are the
trilinear\footnote{These trilinear vertices give rise to Yukawa and the four-scalar
component vertices as in Figure 1. In particular, the four-scalar vertex arises from a
$\Phi\Phi\Phi$ vertex connected to a $\bar\Phi\bar\Phi\bar\Phi$ via a propagator
of an auxiliary $F$ field.}
 ones arising from the superpotential ${\cal W}$ in \eqref{superpotg2} and its
Hermitian conjugate ${\cal \bar{W}}$. These vertices and the superfield propagator are
shown in Figure 9.
\begin{figure}[t]
\label{fig:9beta}
\psfrag{A}{\LARGE${\Phi_1}$}
\psfrag{B}{\LARGE${\Phi_2}$}
\psfrag{C}{\LARGE${\Phi_3}$}
\psfrag{D}{\LARGE${\bar\Phi_1}$}
\psfrag{G}{\LARGE${\bar\Phi_2}$}
\psfrag{H}{\LARGE${\bar\Phi_3}$}
\psfrag{W}{\LARGE${ih \,e^{ i\pi \beta }}$}
\psfrag{X}{\LARGE${-ih \,e^{ -i\pi \beta }}$}
\psfrag{Y}{\LARGE${ih^* \,e^{ i\pi \beta^* }}$}
\psfrag{Z}{\LARGE${-ih^* \,e^{ -i\pi \beta^* }}$}
\psfrag{E}{\LARGE${\Phi_i}$}
\psfrag{F}{\LARGE${\bar\Phi_i}$}
\begin{center}
\scalebox{0.55}{\includegraphics{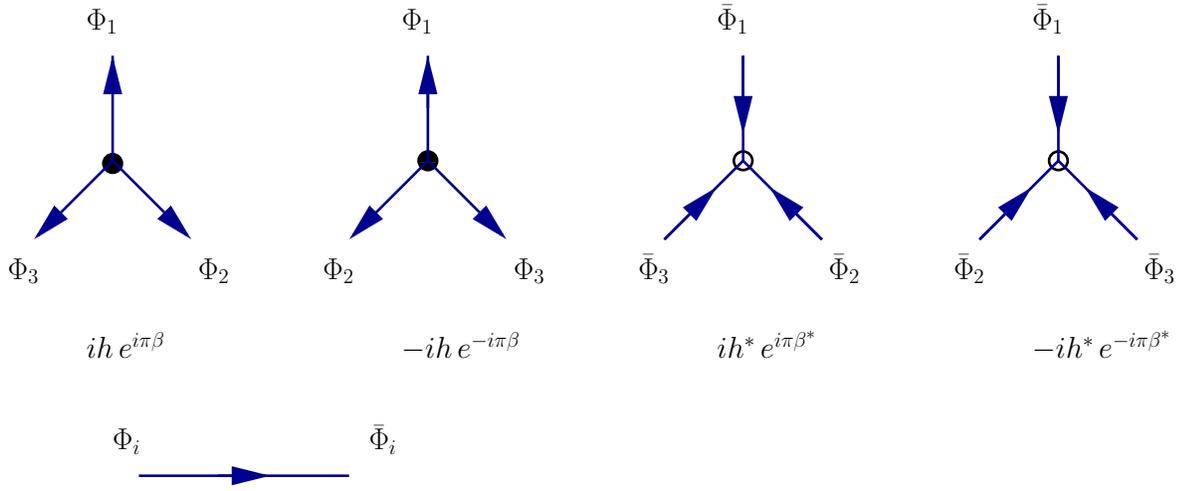}}
\end{center}
\caption{\it Supergraph Feynman rules for (anti)-chiral superfields
corresponding to the superpotential in Eq.~\eqref{superpotg2} and its conjugate.
Arrows are outgoing
for chiral $\Phi$ and incoming for anti-chiral $\bar\Phi$ superfields. The propagator
connects chiral to anti-chiral superfields of the same flavor.
}
\end{figure}

Our goal is to determine the dependence of planar multi-loop amplitudes
on the coupling constant parameters $g$, $h$ and $\beta$ in the deformed theory
and to try to match it with the $g$-dependence of the undeformed $\N=4$
amplitudes using the CFT constraint \eqref{con2l}.
To achieve this we do not need to consider interactions
with the $\cN=1$ vector multiplet, only the matter (anti)-chiral fields,
$\Phi_i$, $\bar\Phi_i$, are relevant.
This is because interactions with vectors
depend only on the original coupling $g$ and not on the
deformation parameters $\beta$ and $h$.
Furthermore, gluons and gluinos $\lambda_4$ and $\bar\lambda_4$
do not change the flavor $i$ of matter superfields $\Phi_i$, $\bar\Phi_i$,
hence they are irrelevant for the matching to $\N=4$.

Hence the relevant
diagrams are the planar vacuum diagrams
built solely from chiral $\Phi_{1,2,3}$ and anti-chiral
$\bar\Phi_{1,2,3}$ matter superfields using super-Feynman rules of Figure 9.
We will call them vacuum $\Phi$-diagrams.
To reproduce the actual amplitude in the full theory,
internal and external vector superfields are supposed to be added in all possible ways
to these effective vacuum $\Phi$-diagrams. However, this
would not affect the overall dependence
on the deformation parameters $h$ and $\beta$.

The simplest vacuum  $\Phi$-diagram with a non-trivial dependence on couplings
occurs at the two-loop level and is shown in Figure 10.
\begin{figure}[h]
\begin{center}
\scalebox{0.55}{\includegraphics{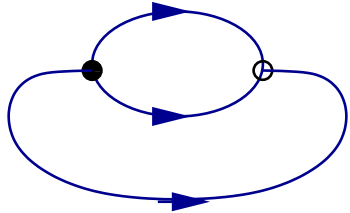}}
\end{center}
\caption{\it Planar vacuum $\Phi$-diagram at the two-loop level.
After summing over all flavors in the loops
the deformation parameters appear in the combination,
$|h|^2 {\rm cosh (2\pi \chi)}=g^2$ via Eq.~\eqref{con2l}.
This matches with the $\N=4$ result.
}
\end{figure}
After summing over all allowed flavors of $\Phi$ circulating in the loops,
one easily finds that the total two-loop contribution is proportional
to  $|h|^2 {\rm cosh (2\pi \chi)}.$ The CFT constraint \eqref{con2l}
guarantees that this matches precisely with the two-loop result in the undeformed
$\cN=4$ theory.

This two-loop matching could have been easily anticipated. The
dependence of the amplitude in
Figure 10 on $h$ and $\beta$ is the same as in the one-loop
$\Phi$-self-energy contributions shown in Figure 11.
\begin{figure}[h]
\label{fig:11beta}
\psfrag{A}{\LARGE${\Phi_1}$}
\psfrag{B}{\LARGE${\Phi_2}$}
\psfrag{C}{\LARGE${\Phi_3}$}
\psfrag{D}{\LARGE${\bar\Phi_1}$}
\psfrag{G}{\LARGE${\bar\Phi_2}$}
\psfrag{H}{\LARGE${\bar\Phi_3}$}
\psfrag{W}{\LARGE${|h|^2 \,e^{ -2\pi \chi }}$}
\psfrag{X}{\LARGE${|h|^2 \,e^{ 2\pi \chi }}$}
\begin{center}
\scalebox{0.55}{\includegraphics{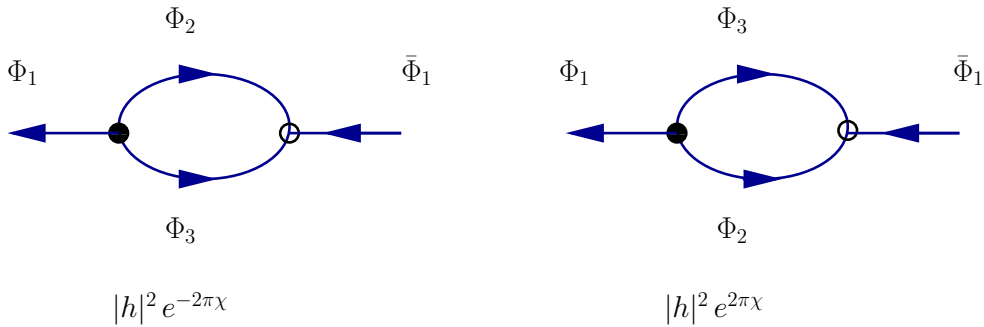}}
\end{center}
\caption{\it Self-energy contributions at one loop involving deformed
supergraph vertices. Total contribution of these two diagrams in the deformed theory
is proportional to $2 |h|^2 {\rm cosh (2\pi \chi)}.$
}
\end{figure}
The latter give rise to the one-loop anomalous dimension $\gamma$
of $\Phi$. The planar CFT constraint \eqref{con2l} arises from
matching $\gamma$ in the deformed theory and to the $\cN=4$
(where it vanishes).
From Figure 11  we read the matching condition as
 $2|h|^2 {\rm cosh} (2\pi \chi)=2g^2$
which is precisely Eq.~\eqref{con2l}.

From this analysis we conclude that we can drop
all self-energy insertions into multi-loop planar
vacuum  $\Phi$-diagrams of the deformed theory -- these insertions
will always match with the $\cN=4$ result.
Hence, we will neglect diagrams which include bubbles as subdiagrams.

It is easy to see that one cannot form triangles using the Feynman rules
in Figure 9. Indeed, each holomorphic in $\Phi$ vertex can be connected
only to an anti-holomorphic one. A triangle would necessarily involve
two adjacent vertices of the same type.
Having lost triangles and bubbles we turn to boxes as building blocks.
The first vacuum diagram involving boxes (with no self-energy contractions)
occurs only at the five-loop level. It is depicted in figure 12.
\begin{figure}[h]
\begin{center}
\scalebox{0.55}{\includegraphics{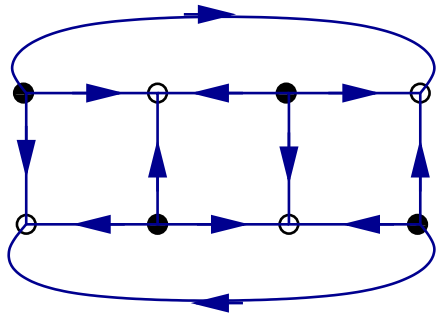}}
\end{center}
\caption{\it
Planar vacuum $\Phi$-diagrams without bubbles
at five-loop level are made of boxes.
}
\end{figure}
Summing over all allowed flavors in the loops in Figure 12 we can find the
total dependence of the five-loop contribution on the deformation parameters.
There is an overall factor of $|h|^8$ but the dependence on $\chi$ is not of the form
${\rm cosh}^4 (2\pi \chi)$. Hence the coupling constant dependence is not given simply by
$g^8$ and this does not match with the five-loop amplitudes in the $\cN=4$
theory.

One has to be a little more careful with this conclusion since the expression
\eqref{con2l} is not an exact solution of the CFT constraint,
and it does in general receive perturbative corrections.
In perturbation theory we can treat of $g^2 \sim |h|^2 \ll 1$ and $\chi \sim 1$.
Then we expect corrections on the right hand side of \eqref{con2l} suppressed by powers
of $g$ or $h.$
One can in principle solve the corrected constraint in terms of $h$ order by order in $g$ and
substitute this solution for $h$ into the bare superpotential \eqref{superpotg2}.
This would amount to fine-tuning the bare deformation order by order in $g$.
It is this fine-tuned
theory which has to be conformally invariant \cite{marginal}.
This fine-tuning is designed to force the anomalous dimensions
and the beta-function vanish (thus removing UV-divergencies
from planar amplitudes) via cancellations between diagrams with different numbers of loops.
It is however unlikely that such cancellations can repair the $\chi$-dependence of the
finite parts of the $n$-point
amplitudes with external gluons.\footnote{
A definitive answer would require an explicit calculation of
all planar five-loop amplitudes with at least four external gluons, which at present is
not feasible.}

%%%%%%%%%%%%%%%%%%%%%%%%%%%%%%%%%%%%%%%%%%%%%%%%%%%%%%%%%%
%%%%%%%%%%%%%%%%%%%%%%%%%%%%%%%%%%%%%%%%%%%%%%%%%%%%%%%%%%

\section{Conclusions} \label{sec:conc}

We have shown that there is a planar equivalence between perturbative amplitudes
in the $\cN=4$ theory and its various deformations.

This equivalence is perturbatively
exact (i) for theories obtained from $\cN=4$ by orbifold projections \cite{BKV};
(ii) for a one parameter family of marginal $\betaR$ deformations considered in
Section {\bf 3} and
(iii) for complex non-supersymmetric deformations considered in Section {\bf 4.1}.
In all these cases, the remarkable iterative structure of planar MHV amplitudes
proposed by Bern, Dixon and Smirnov for the $\cN=4$ theory carries over without
modifications.

For more general marginal deformations studied in Section {\bf 4} the planar equivalence
holds for up to five loops where it (very likely) breaks down.
This fact is intriguing for two reasons: one is that the planar equivalence with the $\cN=4$
theory
holds all the way to four loops. And second, is the fact that even the original $\cN=4$
all-orders proposal of \cite{BDS} was not checked explicitly beyond the three loop level.

Conformal invariance of the theory is likely to play a role in explaining the
underlying principles behind
the iterative structure of MHV amplitudes \cite{BDS}. However on its own conformal
invariance does not appear to be
a sufficient reason for this behavior. It is tempting to speculate that
the transformation properties of the deformed theories
under the $SL(2,Z)$ as well as
the AdS/CFT correspondence will also play a role.

%%%%%%%%%%%%%%%%%%%%%%%%%%%%%%%%%%%%%%%%%%%%%%%%%%%%%%%%%
%%%%%%%%%%%%%%%%%%%%%%%%%%%%%%%%%%%%%%%%%%%%%%%%%%%%%%%%%
\bigskip
\bigskip

\centerline{\bf Acknowledgements}

I would like to thank Lance Dixon for conversations and comments
which greatly improved this paper. I am grateful to Nigel Glover, Simon Badger
and George Georgiou for many useful discussions about amplitudes and loops
and to Adi Armoni for helpful comments
about orientifolding $\cN=4$ SYM.

%%%%%%%%%%%%%%%%%%%%%%%%%%%%%%%%%%%%%%%%%%%%%%%%%%%%%%%%%%%%%%

%%%%%%%%%%%%%%%%%%%%%%%%%%%%%%%%%%%%%%%%%%%%%%%%%%%%%%%%%%%


\newpage
\begin{thebibliography}{99}
{%\small

%\cite{Bern:2005iz}
\bibitem{BDS}
Z.~Bern, L.~J.~Dixon and V.~A.~Smirnov,
%``Iteration of planar amplitudes in maximally supersymmetric Yang-Mills theory at three loops and beyond,''
Phys.\ Rev.\ D {\bf 72} (2005) 085001
[hep-th/0505205].
%%CITATION = HEP-TH 0505205;%%

\bibitem{Witten}
E.~Witten,
%``Perturbative gauge theory as a string theory in twistor space,''
Commun.\ Math.\ Phys.\  {\bf 252} (2004) 189
[hep-th/0312171].
%%CITATION = HEP-TH 0312171;%%

\bibitem{MHVtree}
F.~Cachazo, P.~Svrcek and E.~Witten,
%``MHV vertices and tree amplitudes in gauge theory,''
JHEP {\bf 0409} (2004) 006 [hep-th/0403047] \\
%%CITATION = HEP-TH 0403047;%%
G.~Georgiou and V.~V. Khoze,
%``Tree amplitudes in gauge theory as scalar  {MHV} diagrams,''
{\em JHEP} {\bf 05} (2004) 070
  [hep-th/0404072]\\
%%CITATION = HEP-TH 0404072;%%
C.-J. Zhu,
%``The googly amplitudes in gauge theory,''
{\em JHEP} {\bf 04}  (2004) 032 [hep-th/0403115]\\
%%CITATION = HEP-TH 0403115;%%
J.-B. Wu and C.-J. Zhu,
%``{MHV} vertices and scattering amplitudes in gauge theory,''
{\em JHEP} {\bf 07} (2004) 032
  [hep-th/0406085]\\
%%CITATION = HEP-TH 0406085;%%
I.~Bena, Z.~Bern and D.~A. Kosower,
%``Twistor-space recursive formulation of gauge theory amplitudes,''
{\em Phys. Rev.} {\bf D71} (2005) 045008 [hep-th/0406133]\\
%%CITATION = HEP-TH 0406133;%%
D.~A. Kosower,
%``Next-to-maximal helicity violating amplitudes in gauge theory,''
{\em Phys. Rev.} {\bf D71} (2005) 045007 [hep-th/0406175]\\
%%CITATION = HEP-TH 0406175;%%
G.~Georgiou, E.~W.~N. Glover and V.~V. Khoze,
%``Non-{MHV} tree amplitudes in gauge theory,''
{\em JHEP} {\bf 07} (2004) 048
  [hep-th/0407027]\\
%%CITATION = HEP-TH 0407027;%%
L.~J. Dixon, E.~W.~N. Glover and V.~V. Khoze,
%``{MHV} rules for higgs plus  multi-gluon amplitudes,''
  {\em JHEP} {\bf 12} (2004) 015
  [hep-th/0411092]\\
%%CITATION = HEP-TH 0411092;%%
Z.~Bern, D.~Forde, D.~A. Kosower and P.~Mastrolia,
%``Twistor-inspired construction of electroweak vector boson currents,''
  hep-ph/0412167\\
%%CITATION = HEP-PH 0412167;%%
S.~D.~Badger, E.~W.~N.~Glover and V.~V.~Khoze,
 % ``MHV rules for Higgs plus multi-parton amplitudes,''
  JHEP {\bf 0503}, 023 (2005)
  [hep-th/0412275]\\
  %%CITATION = HEP-TH 0412275;%%
T.~G.~Birthwright, E.~W.~N.~Glover, V.~V.~Khoze and P.~Marquard,
%``Multi-gluon collinear limits from MHV diagrams,''
JHEP {\bf 0505} (2005) 013 [hep-ph/0503063];
%%CITATION = HEP-PH 0503063;%%
%``Collinear limits in QCD from MHV rules,''
JHEP {\bf 0507} (2005) 068 [hep-ph/0505219] \\
%%CITATION = HEP-PH 0505219;%%
K.~Risager,
%``A direct proof of the CSW rules,''
hep-th/0508206 \\
%%CITATION = HEP-TH 0508206;%%
P.~Mansfield,
%``The Lagrangian origin of MHV rules,''
hep-th/0511264.
%%CITATION = HEP-TH 0511264;%%

\bibitem{MHVloop}
F.~Cachazo, P.~Svrcek and E.~Witten,
%``Twistor space structure of one-loop amplitudes in gauge theory,''
JHEP {\bf 0410} (2004) 074
[hep-th/0406177]\\
%%CITATION = HEP-TH 0406177;%%
%``Gauge theory amplitudes in twistor space and holomorphic anomaly,''
JHEP {\bf 0410} (2004) 077
[hep-th/0409245]\\
%%CITATION = HEP-TH 0409245;%%
A.~Brandhuber, B.~J.~Spence and G.~Travaglini,
%``One-loop gauge theory amplitudes in N = 4 super Yang-Mills from MHV
%vertices,''
Nucl.\ Phys.\ B {\bf 706} (2005) 150
[arXiv:hep-th/0407214]\\
%%CITATION = HEP-TH 0407214;%%
F.~Cachazo,
%``Holomorphic anomaly of unitarity cuts and one-loop gauge theory amplitudes,''
hep-th/0410077\\
%%CITATION = HEP-TH 0410077;%%
R.~Britto, F.~Cachazo and B.~Feng,
%``Computing one-loop amplitudes from the holomorphic anomaly of unitarity
%cuts,''
Phys.\ Rev.\ D {\bf 71} (2005) 025012
[hep-th/0410179]\\
%%CITATION = HEP-TH 0410179;%%
Z.~Bern, V.~Del Duca, L.~J.~Dixon and D.~A.~Kosower,
%``All non-maximally-helicity-violating one-loop seven-gluon amplitudes in N = 4
%super-Yang-Mills theory,''
Phys.\ Rev.\ D {\bf 71} (2005) 045006
[hep-th/0410224]\\
%%CITATION = HEP-TH 0410224;%%
C.~Quigley and M.~Rozali,
%``One-loop MHV amplitudes in supersymmetric gauge theories,''
JHEP {\bf 0501} (2005) 053
[hep-th/0410278]\\
%%CITATION = HEP-TH 0410278;%%
J.~Bedford, A.~Brandhuber, B.~J.~Spence and G.~Travaglini,
%``A twistor approach to one-loop amplitudes in N = 1 supersymmetric Yang-Mills
%theory,''
Nucl.\ Phys.\ B {\bf 706} (2005) 100
[hep-th/0410280]\\
%%CITATION = HEP-TH 0410280;%%
S.~J.~Bidder, N.~E.~J.~Bjerrum-Bohr, L.~J.~Dixon and D.~C.~Dunbar,
%``N = 1 supersymmetric one-loop amplitudes and the holomorphic anomaly of
%unitarity cuts,''
Phys.\ Lett.\ B {\bf 606} (2005) 189
[hep-th/0410296]\\
%%CITATION = HEP-TH 0410296;%%
R.~Britto, F.~Cachazo and B.~Feng,
%``Generalized unitarity and one-loop amplitudes in N = 4 super-Yang-Mills,''
Nucl.\ Phys.\ B {\bf 725} (2005) 275
[hep-th/0412103]\\
%%CITATION = HEP-TH 0412103;%%
J.~Bedford, A.~Brandhuber, B.~J.~Spence and G.~Travaglini,
%``Non-supersymmetric loop amplitudes and MHV vertices,''
Nucl.\ Phys.\ B {\bf 712} (2005) 59
[hep-th/0412108]\\
%%CITATION = HEP-TH 0412108;%%
Z.~Bern, L.~J.~Dixon and D.~A.~Kosower,
%``All next-to-maximally helicity-violating one-loop gluon amplitudes in N  = 4
%super-Yang-Mills theory,''
Phys.\ Rev.\ D {\bf 72} (2005) 045014
[hep-th/0412210]\\
S.~J.~Bidder, N.~E.~J.~Bjerrum-Bohr, D.~C.~Dunbar and W.~B.~Perkins,
%``One-loop gluon scattering amplitudes in theories with N < 4
%supersymmetries,''
Phys.\ Lett.\ B {\bf 612} (2005) 75
[hep-th/0502028]\\
%%CITATION = HEP-TH 0502028;%%
%%CITATION = HEP-TH 0412210;%%
R.~Britto, E.~Buchbinder, F.~Cachazo and B.~Feng,
%``One-loop amplitudes of gluons in SQCD,''
Phys.\ Rev.\ D {\bf 72} (2005) 065012
[hep-ph/0503132]\\
R.~Britto, B.~Feng, R.~Roiban, M.~Spradlin and A.~Volovich,
%``All split helicity tree-level gluon amplitudes,''
Phys.\ Rev.\ D {\bf 71} (2005) 105017
[hep-th/0503198]\\
%%CITATION = HEP-TH 0503198;%%
S.~J.~Bidder, D.~C.~Dunbar and W.~B.~Perkins,
%``Supersymmetric Ward identities and NMHV amplitudes involving gluinos,''
JHEP {\bf 0508} (2005) 055
[hep-th/0505249]\\
%%CITATION = HEP-TH 0505249;%%
A.~Brandhuber, S.~McNamara, B.~J.~Spence and G.~Travaglini,
%``Loop amplitudes in pure Yang-Mills from generalised unitarity,''
JHEP {\bf 0510} (2005) 011
[hep-th/0506068]\\
%%CITATION = HEP-TH 0506068;%%
E.~I.~Buchbinder and F.~Cachazo,
%``Two-loop amplitudes of gluons and octa-cuts in N = 4 super Yang-Mills,''
JHEP {\bf 0511} (2005) 036
[arXiv:hep-th/0506126]\\
%%CITATION = HEP-TH 0506126;%%
A.~Brandhuber, B.~Spence and G.~Travaglini,
%``From trees to loops and back,''
hep-th/0510253.
%%CITATION = HEP-TH 0510253;%%


\bibitem{RECtree}
R.~Britto, F.~Cachazo and B.~Feng,
%``New recursion relations for tree amplitudes of gluons,''
Nucl.\ Phys.\ B {\bf 715} (2005) 499
[hep-th/0412308]\\
R.~Roiban, M.~Spradlin and A.~Volovich,
%``Dissolving N = 4 loop amplitudes into QCD tree amplitudes,''
Phys.\ Rev.\ Lett.\  {\bf 94}, 102002 (2005)
[hep-th/0412265]\\
%%CITATION = HEP-TH 0412265;%%
R.~Britto, F.~Cachazo, B.~Feng and E.~Witten,
%``Direct proof of tree-level recursion relation in Yang-Mills theory,''
Phys.\ Rev.\ Lett.\  {\bf 94} (2005) 181602
[hep-th/0501052]\\
%%CITATION = HEP-TH 0501052;%%
M.~x.~Luo and C.~k.~Wen,
%``Recursion relations for tree amplitudes in super gauge theories,''
JHEP {\bf 0503} (2005) 004
[hep-th/0501121];
%%CITATION = HEP-TH 0501121;%%
%``Compact formulas for all tree amplitudes of six partons,''
Phys.\ Rev.\ D {\bf 71} (2005) 091501
[hep-th/0502009]\\
%%CITATION = HEP-TH 0502009;%%
S.~D.~Badger, E.~W.~N.~Glover, V.~V.~Khoze and P.~Svrcek,
%``Recursion relations for gauge theory amplitudes with massive particles,''
JHEP {\bf 0507} (2005) 025
[hep-th/0504159]\\
%%CITATION = HEP-TH 0504159;%%
S.~D.~Badger, E.~W.~N.~Glover and V.~V.~Khoze,
%``Recursion relations for gauge theory amplitudes with massive vector bosons
%and fermions,''
[hep-th/0507161].
%%CITATION = HEP-TH 0507161;%%

\bibitem{RECloop}
Z.~Bern, L.~J.~Dixon and D.~A.~Kosower,
%``On-shell recurrence relations for one-loop QCD amplitudes,''
Phys.\ Rev.\ D {\bf 71} (2005) 105013
[hep-th/0501240];
%%CITATION = HEP-TH 0501240;%%
%``The last of the finite loop amplitudes in QCD,''
Phys.\ Rev.\ D {\bf 72} (2005) 125003
[hep-ph/0505055];
%%CITATION = HEP-PH 0505055;%%
%``Bootstrapping multi-parton loop amplitudes in QCD,''
hep-ph/0507005\\
%%CITATION = HEP-PH 0507005;%%
Z.~Bern, N.~E.~J.~Bjerrum-Bohr, D.~C.~Dunbar and H.~Ita,
%``Recursive calculation of one-loop QCD integral coefficients,''
hep-ph/0507019.
%%CITATION = HEP-PH 0507019;%%

\bibitem{QMconf}
{\it From Twistors to Amplitudes}, QMUL Workshop, 3-5 November 2005, on-line talks at
{\tt http://www.strings.ph.qmul.ac.uk/{\%}7Eandreas/FTTA/program.htm}

%\cite{Leigh:1995ep}
\bibitem{LS}
R.~G.~Leigh and M.~J.~Strassler,
%``Exactly marginal operators and duality in four-dimensional N=1 supersymmetric gauge theory,''
Nucl.\ Phys.\ B {\bf 447}, 95 (1995)
[hep-th/9503121].
%%CITATION = HEP-TH 9503121;%%

\bibitem{marginal}
  A.~Parkes and P.~C.~West,
  %``Finiteness In Rigid Supersymmetric Theories,''
  Phys.\ Lett.\ B {\bf 138}, 99 (1984)\\
  %%CITATION = PHLTA,B138,99;%%
  P.~C.~West,
  %``The Yukawa Beta Function In N=1 Rigid Supersymmetric Theories,''
  Phys.\ Lett.\ B {\bf 137}, 371 (1984)\\
  %%CITATION = PHLTA,B137,371;%%
   D.~R.~T.~Jones and L.~Mezincescu,
  %``The Chiral Anomaly And A Class Of Two Loop Finite Supersymmetric Gauge
  %Theories,''
  Phys.\ Lett.\ B {\bf 138}, 293 (1984)\\
  %%CITATION = PHLTA,B138,293;%%
 S.~Hamidi, J.~Patera and J.~H.~Schwarz,
  %``Chiral Two Loop Finite Supersymmetric Theories,''
  Phys.\ Lett.\ B {\bf 141}, 349 (1984)\\
  %%CITATION = PHLTA,B141,349;%%
 W.~Lucha and H.~Neufeld,
  %``Finiteness Of Quantum Field Theories And Supersymmetry,''
  Phys.\ Lett.\ B {\bf 174}, 186 (1986)\\
  %%CITATION = PHLTA,B174,186;%%
 D.~R.~T.~Jones,
  %``Coupling Constant Reparametrization And Finite Field Theories,''
  Nucl.\ Phys.\ B {\bf 277}, 153 (1986)\\
  %%CITATION = NUPHA,B277,153;%%
   X.~d.~Jiang and X.~J.~Zhou,
  %``Two Loop Finite Supersymmetric Theories Of SU(N),''
  Commun.\ Theor.\ Phys.\  {\bf 5}, 179 (1986);
  %%CITATION = CTPMD,5,179;%%
%``A Criterion For Existence Of Finite To All Orders N=1 Sym Theories,''
  Phys.\ Rev.\ D {\bf 42}, 2109 (1990)\\
  %%CITATION = PHRVA,D42,2109;%%
  A.~V.~Ermushev, D.~I.~Kazakov and O.~V.~Tarasov,
  %``Finite N=1 Supersymmetric Grand Unified Theories,''
  Nucl.\ Phys.\ B {\bf 281}, 72 (1987)\\
  %%CITATION = NUPHA,B281,72;%%
%
  D.~I.~Kazakov,
  %``Finite N=1 Susy Gauge Field Theories,''
  Mod.\ Phys.\ Lett.\ A {\bf 2}, 663 (1987).
  %%CITATION = MPLAE,A2,663;%%

%\cite{Lunin:2005jy}
\bibitem{LM}
O.~Lunin and J.~Maldacena,
%``Deforming field theories with U(1) x U(1) global symmetry and their gravity duals,''
JHEP {\bf 0505} (2005) 033
[hep-th/0502086].
%%CITATION = HEP-TH 0502086;%%

\bibitem{BKV}
M.~Bershadsky, Z.~Kakushadze and C.~Vafa,
%``String expansion as large N expansion of gauge theories,''
Nucl.\ Phys.\ B {\bf 523} (1998) 59
[hep-th/9803076]\\
%%CITATION = HEP-TH 9803076;%%
M.~Bershadsky and A.~Johansen,
%``Large N limit of orbifold field theories,''
Nucl.\ Phys.\ B {\bf 536} (1998) 141
{hep-th/9803249].
%%CITATION = HEP-TH 9803249;%%

\bibitem{BDDK}
Z.~Bern, L.~J.~Dixon, D.~C.~Dunbar and D.~A.~Kosower,
%``One loop n point gauge theory amplitudes, unitarity and collinear limits,''
Nucl.\ Phys.\ B {\bf 425} (1994) 217
[hep-ph/9403226];
%%CITATION = HEP-PH 9403226;%%
%``Fusing gauge theory tree amplitudes into loop amplitudes,''
Nucl.\ Phys.\ B {\bf 435} (1995) 59
[hep-ph/9409265].
%%CITATION = HEP-PH 9409265;%%

\bibitem{BRY}
Z.~Bern, J.~S.~Rozowsky and B.~Yan,
%``Two-loop four-gluon amplitudes in N = 4 super-Yang-Mills,''
Phys.\ Lett.\ B {\bf 401} (1997) 273
[hep-ph/9702424]\\
%%CITATION = HEP-PH 9702424;%%
Z.~Bern, L.~J.~Dixon, D.~C.~Dunbar, M.~Perelstein and J.~S.~Rozowsky,
%``On the relationship between Yang-Mills theory and gravity and its
%implication for ultraviolet divergences,''
Nucl.\ Phys.\ B {\bf 530} (1998) 401
[hep-th/9802162].
%%CITATION = HEP-TH 9802162;%%

\bibitem{ABDK}
C.~Anastasiou, Z.~Bern, L.~J.~Dixon and D.~A.~Kosower,
%``Planar amplitudes in maximally supersymmetric Yang-Mills theory,''
Phys.\ Rev.\ Lett.\  {\bf 91} (2003) 251602
[hep-th/0309040];
%%CITATION = HEP-TH 0309040;%%
%``Cross-order relations in N = 4 supersymmetric gauge theories,''
[hep-th/0402053].
%%CITATION = HEP-TH 0402053;%%

\bibitem{Smirnov}
V.~A.~Smirnov,
%``Analytical result for dimensionally regularized massless on-shell  planar
%triple box,''
Phys.\ Lett.\ B {\bf 567} (2003) 193
[hep-ph/0305142].
%%CITATION = HEP-PH 0305142;%%

\bibitem{STY}
G.~Sterman and M.~E.~Tejeda-Yeomans,
%``Multi-loop amplitudes and resummation,''
Phys.\ Lett.\ B {\bf 552} (2003) 48
[hep-ph/0210130].
%%CITATION = HEP-PH 0210130;%%

\bibitem{MS}
L.~Magnea and G.~Sterman,
%``Analytic Continuation Of The Sudakov Form-Factor In QCD,''
Phys.\ Rev.\ D {\bf 42} (1990) 4222.
%%CITATION = PHRVA,D42,4222;%%

\bibitem{Catani}
S.~Catani,
%``The singular behaviour of {QCD} amplitudes at two-loop order,''
Phys.\ Lett.\ B {\bf 427} (1998) 161
[hep-ph/9802439].
%%CITATION = HEP-PH 9802439;%%


\bibitem{AdSCFT}
J.~M.~Maldacena,
%``The large N limit of superconformal field theories and supergravity,''
Adv.\ Theor.\ Math.\ Phys.\  {\bf 2} (1998) 231
[Int.\ J.\ Theor.\ Phys.\  {\bf 38} (1999) 1113]
[hep-th/9711200]\\
%%CITATION = HEP-TH 9711200;%
S.~S.~Gubser, I.~R.~Klebanov and A.~M.~Polyakov,
%``Gauge theory correlators from non-critical string theory,''
Phys.\ Lett.\ B {\bf 428} (1998) 105
[hep-th/9802109]\\
%%CITATION = HEP-TH 9802109;%%
E.~Witten,
%``Anti-de Sitter space and holography,''
Adv.\ Theor.\ Math.\ Phys.\  {\bf 2} (1998) 253
[hep-th/9802150].
%%CITATION = HEP-TH 9802150;%%



\bibitem{AdSCFT2}
D.~Berenstein, J.~M.~Maldacena and H.~Nastase,
%``Strings in flat space and pp waves from N = 4 super Yang Mills,''
JHEP {\bf 0204} (2002) 013
[hep-th/0202021]\\
%%CITATION = HEP-TH 0202021;%%
S.~S.~Gubser, I.~R.~Klebanov and A.~M.~Polyakov,
%``A semi-classical limit of the gauge/string correspondence,''
Nucl.\ Phys.\ B {\bf 636} (2002) 99
[hep-th/0204051].
%%CITATION = HEP-TH 0204051;%%


\bibitem{Strassler}
M.~J.~Strassler,
%``On methods for extracting exact non-perturbative results in
%non-supersymmetric gauge theories,''
[hep-th/0104032]\\
%%CITATION = HEP-TH 0104032;%%
D.~Tong,
%``Comments on condensates in non-supersymmetric orbifold field theories,''
JHEP {\bf 0303} (2003) 022
[hep-th/0212235]\\
%%CITATION = HEP-TH 0212235;%%
P.~Kovtun, M.~Unsal and L.~G.~Yaffe,
%``Non-perturbative equivalences among large N(c) gauge theories with  adjoint
%and bifundamental matter fields,''
JHEP {\bf 0312} (2003) 034
[hep-th/0311098].
%%CITATION = HEP-TH 0311098;%%

\bibitem{ASV}
A.~Armoni, M.~Shifman and G.~Veneziano,
%``Exact results in non-supersymmetric large N orientifold field theories,''
Nucl.\ Phys.\ B {\bf 667} (2003) 170
[hep-th/0302163];
%%CITATION = HEP-TH 0302163;%%
%``From super-Yang-Mills theory to QCD: Planar equivalence and its
%implications,''
[hep-th/0403071].
%%CITATION = HEP-TH 0403071;%%

%\cite{Freedman:2005cg}
\bibitem{FG}
D.~Z.~Freedman and U.~Gursoy,
%``Comments on the beta-deformed N = 4 SYM theory,''
JHEP {\bf 0511} (2005) 042
[hep-th/0506128].
%%CITATION = HEP-TH 0506128;%%

%\cite{Penati:2005hp}
\bibitem{PSZ}
S.~Penati, A.~Santambrogio and D.~Zanon,
%``Two-point correlators in the beta-deformed N = 4 SYM at the next-to-leading
%order,''
JHEP {\bf 0510} (2005) 023
[hep-th/0506150].
%%CITATION = HEP-TH 0506150;%%

%\cite{Mauri:2005pa}
\bibitem{MPSZ}
A.~Mauri, S.~Penati, A.~Santambrogio and D.~Zanon,
%``Exact results in planar N = 1 superconformal Yang-Mills theory,''
JHEP {\bf 0511} (2005) 024
[hep-th/0507282].
%%CITATION = HEP-TH 0507282;%%

\bibitem{NSVZ}
V.~A.~Novikov, M.~A.~Shifman, A.~I.~Vainshtein and V.~I.~Zakharov,
%``Beta Function In Supersymmetric Gauge Theories: Instantons Versus Traditional Approach,''
Phys.\ Lett.\ B {\bf 166} (1986) 329; \\
%%CITATION = PHLTA,B166,329;%%
%
M.~A.~Shifman and A.~I.~Vainshtein,
%``Solution Of The Anomaly Puzzle In Susy Gauge Theories And The Wilson Operator Expansion,''
Nucl.\ Phys.\ B {\bf 277} (1986) 456;
%%CITATION = NUPHA,B277,456;%%
%
%M.~A.~Shifman and A.~I.~Vainshtein,
%``On holomorphic dependence and infrared effects in supersymmetric gauge theories,''
Nucl.\ Phys.\ B {\bf 359} (1991) 571.
%%CITATION = NUPHA,B359,571;%%

%\cite{Kulaxizi:2004pa}
\bibitem{KZ}
M.~Kulaxizi and K.~Zoubos,
%``Marginal deformations of N = 4 SYM from open / closed twistor strings,''
[hep-th/0410122].
%%CITATION = HEP-TH 0410122;%%



%\cite{Filk:1996dm}
\bibitem{Filk}
T.~Filk,
%``Divergencies in a field theory on quantum space,''
Phys.\ Lett.\ B {\bf 376} (1996) 53.
%%CITATION = PHLTA,B376,53;%%

%\cite{Ishibashi:1999hs}
\bibitem{IIKK}
N.~Ishibashi, S.~Iso, H.~Kawai and Y.~Kitazawa,
%``Wilson loops in noncommutative Yang-Mills,''
Nucl.\ Phys.\ B {\bf 573} (2000) 573
[hep-th/9910004].
%%CITATION = HEP-TH 9910004;%%


%\cite{Minwalla:1999px}
\bibitem{Minwalla}
S.~Minwalla, M.~Van Raamsdonk and N.~Seiberg,
%``Noncommutative perturbative dynamics,''
JHEP {\bf 0002} (2000) 020
[hep-th/9912072].
%%CITATION = HEP-TH 9912072;%%

%\cite{Dymarsky:2005uh}
\bibitem{DKR}
A.~Dymarsky, I.~R.~Klebanov and R.~Roiban,
%``Perturbative search for fixed lines in large N gauge theories,''
JHEP {\bf 0508} (2005) 011
[hep-th/0505099]; 
%%CITATION = HEP-TH 0505099;%%
JHEP {\bf 0511} (2005) 038
[hep-th/0509132].
%%CITATION = HEP-TH 0509132;%%

}}

\end{thebibliography}
\end{document}